%% file: mor04.tex
\documentclass[11pt]{article}

\usepackage{moriond,epsfig,relsize}  
\usepackage{multirow}

\input babarsym
\input Definitions

\newcommand{\ubma}[1]{\unboldmath{$#1$}}
\def\Abar    {\kern 0.18em\overline{\kern -0.18em A}{} }

\renewcommand{\retapip}{\ensuremath{5.3\pm1.0\pm0.3}}
\renewcommand{\Aetapip}{\ensuremath{-0.44\pm0.18\pm0.01}}
\renewcommand{\setapip}{\ensuremath{7.9}}
\renewcommand{\retaKp}{\ensuremath{3.4\pm0.8\pm0.2}}
\renewcommand{\AetaKp}{\ensuremath{-0.52\pm0.24\pm0.01}}
\renewcommand{\setaKp}{\ensuremath{6.1}}
\renewcommand{\retaKz}{\ensuremath{2.9\pm1.0\pm 0.2}}
\renewcommand{\uletaKz}{\ensuremath{5.2}}
\renewcommand{\setaKz}{\ensuremath{3.3}}

\renewcommand{\retappip}{\ensuremath{2.7\pm1.2\pm 0.3}}
\renewcommand{\uletappip}{\ensuremath{4.5}}

\renewcommand{\setappip}{\ensuremath{3.4}}

\renewcommand{\retapiz}{\ensuremath{0.7^{+1.1}_{-0.9}\pm 0.3}}
\renewcommand{\uletapiz}{\ensuremath{2.5}}
\renewcommand{\setapiz}{\ensuremath{0.8}}			

\renewcommand{\retaKstp}{\ensuremath{25.6\pm4.0\pm2.4}}
\renewcommand{\AetaKstp}{\ensuremath{+0.13\pm0.14\pm0.02}}
\renewcommand{\setaKstp}{\ensuremath{9}}

\renewcommand{\retaKstz}{\ensuremath{18.6\pm2.3\pm 1.2}}
\renewcommand{\AetaKstz}{\ensuremath{+0.02\pm0.11\pm0.02}}
\renewcommand{\setaKstz}{\ensuremath{11}}

\renewcommand{\retarhop}{\ensuremath{9.2\pm3.4\pm1.0}}
\renewcommand{\uletarhop}{\ensuremath{14}}
\renewcommand{\setarhop}{\ensuremath{3.5}}

\renewcommand{\retarhoz}{\ensuremath{-1.1^{+0.7}_{-0.9}\pm0.4}}      
\renewcommand{\uletarhoz}{\ensuremath{1.5}}
\renewcommand{\setarhoz}{\ensuremath{-}}                        

\renewcommand{\retappiz}{\ensuremath{1.0^{+1.4}_{-1.0} \pm 0.8}}
\renewcommand{\uletappiz}{\ensuremath{3.7}}
\renewcommand{\setappiz}{\ensuremath{0.7}}

\renewcommand{\retapKstp}{\ensuremath{6.3^{+4.6}_{-3.6}\pm 1.8}}
\renewcommand{\uletapKstp}{\ensuremath{14}}
\renewcommand{\setapKstp}{\ensuremath{1.9}}

\renewcommand{\retapKstz}{\ensuremath{4.1^{+2.1}_{-1.8}\pm 1.2}}
\renewcommand{\uletapKstz}{\ensuremath{7.6}}
\renewcommand{\setapKstz}{\ensuremath{2.1}}

\renewcommand{\retaprhop}{\ensuremath{12.9^{+6.2}_{-5.5}\pm 2.0}}
\renewcommand{\uletaprhop}{\ensuremath{22}}
\renewcommand{\setaprhop}{\ensuremath{2.6}}

\renewcommand{\retaprhoz}{\ensuremath{0.8^{+1.7}_{-1.2}\pm 0.9}}      
\renewcommand{\uletaprhoz}{\ensuremath{4.3}}
\renewcommand{\setaprhoz}{\ensuremath{0.5}}                        

\renewcommand{\romegapiz}{\ensuremath{-0.6^{+0.7}_{-0.5}\pm0.2}}
\renewcommand{\ulomegapiz}{\ensuremath{1.2}}

\renewcommand{\rphipiz}{\ensuremath{0.2^{+0.4}_{-0.3}\pm0.1}}
\renewcommand{\ulphipiz}{\ensuremath{1.0}}
\renewcommand{\sphipiz}{\ensuremath{0.7}}

\newcommand{\fetaeta}{\ensuremath{\eta\eta}}
\newcommand{\fetaphi}{\ensuremath{\eta\phi}}
\newcommand{\fetaomega}{\ensuremath{\eta\omega}}
\newcommand{\fetaetap}{\ensuremath{\eta\etapr}}
\newcommand{\fetapetap}{\ensuremath{\etapr\etapr}}
\newcommand{\fetapomega}{\ensuremath{\etapr \omega}}
\newcommand{\fetapphi}{\ensuremath{\etapr\phi}}
\newcommand{\fphiphi}{\ensuremath{\phi\phi}}
\newcommand{\etaeta}{\ensuremath{\Bz\ra\fetaeta}}
\newcommand{\etaphi}{\ensuremath{\Bz\ra\fetaphi}}
\newcommand{\etaomega}{\ensuremath{\Bz\ra\fetaomega}}
\newcommand{\etaetap}{\ensuremath{\Bz\ra\fetaetap}}
\newcommand{\etapetap}{\ensuremath{\Bz\ra\fetapetap}}
\newcommand{\etapomega}{\ensuremath{\Bz\ra\fetapomega}}
\newcommand{\etapphi}{\ensuremath{\Bz\ra\fetapphi}}
\newcommand{\phiphi}{\ensuremath{\Bz\ra\fphiphi}}
\newcommand{\retaomega}{\ensuremath{4.0^{+1.3}_{-1.2}\pm0.4}}

\newcommand{\retaetap}{\ensuremath{0.6^{+2.1}_{-1.7}\pm 1.1}}
\newcommand{\uletaetap}{\ensuremath{4.6}}
\newcommand{\rphiphi}{\ensuremath{0.3^{+0.7}_{-0.4}\pm0.1}}

\renewcommand{\razerompip}{\ensuremath{2.8^{+1.5}_{-1.3}\pm 0.7}}
\renewcommand{\ulazerompip}{\ensuremath{5.1}\xspace}
\renewcommand{\sazerompip}{\ensuremath{2.0}\xspace}
\renewcommand{\razeromKp}{\ensuremath{0.4^{+1.0}_{-0.8}\pm 0.2}}
\renewcommand{\ulazeromKp}{\ensuremath{2.1}\xspace}
\renewcommand{\sazeromKp}{\ensuremath{0.4}\xspace}
\renewcommand{\razeromKz}{\ensuremath{-1.5^{+2.4}_{-1.8}\pm 0.8}}
\renewcommand{\ulazeromKz}{\ensuremath{3.9}\xspace}
\renewcommand{\sazeromKz}{\ensuremath{0.6}\xspace}
\renewcommand{\razerozpip}{\ensuremath{3.6^{+2.1}_{-1.9}\pm 0.8}}
\renewcommand{\ulazerozpip}{\ensuremath{6.7}\xspace}
\renewcommand{\sazerozpip}{\ensuremath{1.9}\xspace}
\renewcommand{\razerozKp}{\ensuremath{-3.7^{+1.6}_{-1.3}\pm 0.5}}
\renewcommand{\ulazerozKp}{\ensuremath{1.8}\xspace}
\renewcommand{\sazerozKp}{\ensuremath{0.0}\xspace}
\renewcommand{\razerozKz}{\ensuremath{2.8^{+3.1}_{-2.4}\pm 1.1}}
\renewcommand{\ulazerozKz}{\ensuremath{7.8}\xspace}
\renewcommand{\sazerozKz}{\ensuremath{1.0}\xspace}

\newcommand{\fphikst}{\ensuremath{\phi\Kstarz}}
\newcommand{\phikst}{\ensuremath{\Bz\ra\fphikst}}

\begin{document}

\begin{flushleft}
\babar-PROC-04/021\\
\end{flushleft}
\par\vskip .5cm

\title{Recent Results in Charmless Hadronic B Decays from BABAR}

\author{ J. G. Smith \\(Representing the BABAR Collaboration)}

\address{Department of Physics, University of Colorado\\
Boulder, CO 80309, USA}

\maketitle\abstracts{
We report results from five analyses based on data taken with the \babar\
detector at the PEP-II asymmetric $e^+e^-$ collider.  Included are
branching fraction measurements for many $B$-meson decays involving $\eta$, 
\etapr, $\omega$, $\phi$ or \azero\ mesons and the final state $\KS\pip\pim$,
and a full angular analysis of the decay \phikst.
}
  
\section{Introduction}
Many interesting new results from \babar\ for charmless hadronic $B$ decays 
were presented previously at the 
Electroweak session of the XXXIXth Rencontres de Moriond.  For new measurements 
of \stwob\ from four final states ($\phi\Kz$, $\Kp\Km\KS$, $\piz\KS$, 
$f_0\KS$), see the writeup by Marc Verderi.  Also a new preliminary result for 
the decay $\Bz\to\rhop\rhom$, with a measurement of the CKM angle
$\alpha$ was presented in a talk by Lydia Roos.  Finally a measurement of
the time-dependent asymmetry of the decay $\Bz\to\piz\KS\gamma$ was shown
by Eugenio Paoloni.  With adequate data, the latter mode can provide
interesting constraints on new physics.

In this paper I will report on five other new analyses of charmless
hadronic $B$ decays.  The first involves \B\ decays to $\etaprp K^*$,
$\etaprp\rho$, $\etaprp\pi^0$, $\omega\pi^0$, and $\phi\pi^0$.\cite{PRD}
Substantial signals are seen for \etaKst\ and limits are provided for the
other modes.  The decay \etapKst\ is particularly interesting since it
provides limits on a flavor-singlet amplitude.\cite{chiang,beneke}  The
second analysis searches for eight isoscalar final states.\cite{isoscalar}
In addition to the interest in observing signals should the branching
fractions be large enough, these channels are interesting because they 
can provide constraints on the expected value of \stwob\ for the modes
\etapKz\ and $\Bz\to\phi\Kz$.\cite{grossman,GRZ}  These channels provide
constraints on the size of the color-suppressed tree amplitudes for these
penguin-dominated channels.  The third analysis involves a search for $B$
decays to the scalar \azero\ meson accompanied by pions or kaons.  Little
is known about decays involving scalars.  The fourth analysis is a 
fairly precise measurement of the decay $B\to\KS\pip\pim$.  The last
analysis measures the polarization and potential $CP$-violating terms in
the full angular analysis of the decay $B\to\phi \Kstz$.  

\section{Datasets and analysis details}
The results presented here are based on data collected with the \babar\ 
detector~\cite{BABARNIM} at the PEP-II asymmetric $e^+e^-$ collider
located at the Stanford Linear Accelerator Center.  Most analyses use a
sample of 89 million \BB\ pairs, recorded at the $\Upsilon (4S)$
resonance (center-of-mass energy $\sqrt{s}=10.58\ \gev$).  The $B\to\phi
\Kstz$ analysis uses a sample of 124 million \BB\ pairs.

A $B$-meson candidate is characterized kinematically by the
energy-substituted mass \mes\ and by the energy difference \DE, defined as
\begin{eqnarray}
\mes &=& \sqrt{{1\over4}s - \pvec_B^{*2}} \qquad \mbox{and}  \\
\DE  &=& E_B^*-\half\sqrt{s} \ ,
\end{eqnarray}
where $(E_B,\pvec_B)$ is the $B$-candidate four vector and $s$ is the
square of the invariant mass of the electron-positron system; the asterisk 
denotes the value in the \UfourS\ frame.  All analyses use these two quantities 
in unbinned maximum-likelihood fits which also have invariant masses of 
quasi-two-body resonances in the final states and a Fisher discriminant
that is sensitive to event shape.

\section{Measurements of $\etaprp K^*$ and related decays}
\begin{figure}[htb!]
\begin{center}
 \scalebox{0.7}{
  \includegraphics[width=\linewidth]{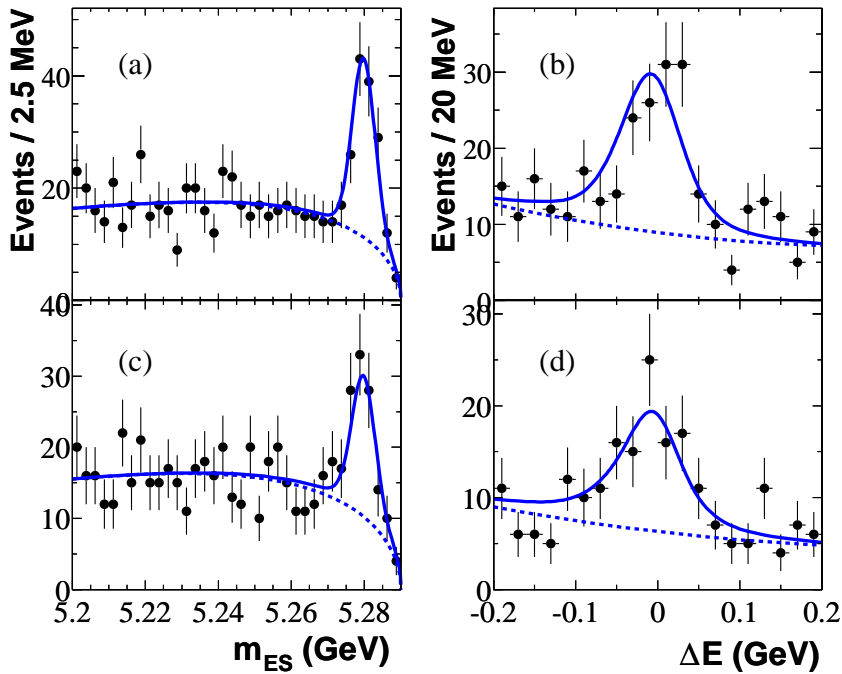}
  }
\end{center}
 \caption{\label{fig:etakstproj} Projections of the $B$-candidate
 \mes\ and \DE\ distributions for (a),(b) \etaKstz\ and (c),(d) \etaKstp.
 Not all events are shown since these plots are made with a requirement on 
 the likelihood.
 }
\end{figure}

We have searched for the \B\ decays to $\etaprp K^*$,
$\etaprp\rho$, $\etaprp\pi^0$, $\omega\pi^0$, and $\phi\pi^0$.  We find a
substantial signal for both charge states of the \etaKst\ decay as shown
in the projection plots in Fig. \ref{fig:etakstproj}.
These results are tabulated in Table \ref{tab:restab_prd} along with 
previous results for the $\etaprp K$ and $\etaprp\pi$ decays.  Thus we
have completed the measurement of the four $(\eta,\etapr)(K,\Kstar)$ final
states with a sensitivity in the branching fraction of a few times $10^{-6}$.
We find no significant signal for \etapKst; the 90\% C.L. upper limit
is not yet precise enough to determine whether a flavor-singlet component
is present for this decay, though we do restrict the size of such a
contribution.  See Ref.~\ref{chiang} and references therein for a discussion
of this issue.  We also have evidence for the decay \etarhop\ with a
significance of 3.5$\sigma$.

\begin{table*}[ht]
\caption{We show the significance ${\cal S}(\sigma)$ (including systematic
errors), fit branching fractions \calB, 90\% C.L. upper limits, and charge 
asymmetries for the 12 new measurements as well as six related measurements 
(above the line) that were published recently.\protect\cite{etaprPRL,etaomegaPRL}}
\label{tab:restab_prd}
\begin{center}
\begin{tabular}{lrccc}
    \dbline
Mode          		&~${\cal S}(\sigma$)& ${\cal B} (10^{-6})$ & UL $(10^{-6})$ & \acp  \\
\sgline
\ubma{\etapKp}		&\ubma{>10}	& \ubma{76.9\pm3.5} &  & \ubma{0.037\pm0.045} \\
\ubma{\etapKz}		&\ubma{>10}	& \ubma{60.6\pm5.6}&  &	\\
\ubma{\etapip}           &\ubma{\setapip} &\ubma{\retapip}& &\ubma{\Aetapip} \\
\ubma{\etaKp}            &\ubma{\setaKp}  &\ubma{\retaKp} & &\ubma{\AetaKp} \\
\ubma{\etaKz}            &\ubma{\setaKz}  &\ubma{\retaKz} &\ubma{<\uletaKz} &    \\
\ubma{\etappip}          &\ubma{\setappip}&\ubma{\retappip}&\ubma{<\uletappip} &    \\
\sgline
\ubma{\etaKstp}		&\ubma{\setaKstp~}&~\ubma{\retaKstp}~& &~  \ubma{\AetaKstp}\\
\ubma{\etaKstz}		&\ubma{\setaKstz~}&~\ubma{\retaKstz}~&&~  \ubma{\AetaKstz}\\
\ubma{\etarhop}		&\ubma{\setarhop}&\ubma{\retarhop}&\ubma{<\uletarhop}&                         \\
\ubma{\etarhoz}		&\ubma{\setarhoz~}&\ubma{\retarhoz}&\ubma{<\uletarhoz}&    			\\
\ubma{\etapiz}		&\ubma{\setapiz}	 &\ubma{\retapiz}&\ubma{<\uletapiz}&   			\\
\ubma{\etapKstp}		&\ubma{\setapKstp}&\ubma{\retapKstp}&\ubma{<\uletapKstp}&   \\
\ubma{\etapKstz}		&\ubma{\setapKstz}&\ubma{\retapKstz}&\ubma{<\uletapKstz}&  \\
\ubma{\etaprhop}		&\ubma{\setaprhop}&\ubma{\retaprhop}&\ubma{<\uletaprhop}&   \\
\ubma{\etaprhoz}		&\ubma{\setaprhoz}&\ubma{\retaprhoz}&\ubma{<\uletaprhoz}&  \\
\ubma{\etappiz}		&\ubma{\setappiz}&\ubma{\retappiz}&\ubma{<\uletappiz}&   \\
\ubma{\omegapiz}		&\ubma{-~}    	&\ubma{\romegapiz}&\ubma{<\ulomegapiz}&  	\\
\ubma{\phipiz}		&\ubma{\sphipiz}	&\ubma{\rphipiz}&\ubma{<\ulphipiz}&   	\\
\dbline
\end{tabular}
\end{center}
\end{table*}

\section{Search for isoscalar charmless decays}
We have searched for eight isoscalar charmless decays.  These decays are 
particularly interesting because they
can provide constraints on the expected value of \stwob\ for the modes
\etapKz\ and $\Bz\to\phi\Kz$.\cite{grossman,GRZ}
Results are summarized in Table \ref{tab:res_isoscalar}.  The 4.3$\sigma$
signal in $\Bz\to\eta\omega$ is unexpected and may be a fluctuation; more
data will be required to see if this is interesting.  The limits on all
of these modes have improved the understanding of the expected value of 
\stwob\ for \etapKz\ so that the model-independent precision is now 0.10.
\cite{GRZ}  This is an improvement of about a factor of five on the
previous limits.\cite{grossman}

\begin{table*}[htb!]
\caption{
Significance ${\cal S}(\sigma)$ (including systematic uncertainties), measured branching
fraction \calB, and 90\% C.L. upper limits (UL) from this and previous
measurements by CLEO.
}
\label{tab:res_isoscalar}
\begin{center}
\begin{tabular}{lcccc}
\dbline
Mode&\quad S($\sigma$) \quad &\quad \bfemsix
\quad&\quad UL $(10^{-6})$ \quad &\quad CLEO UL $(10^{-6})$ \cite{CLEO} \quad \\
\tbline
\ubma{\etaeta}    &\ubma{0.0}&\ubma{-0.9^{+1.6}_{-1.4}\pm 0.7} &\ubma{~ < 2.8} & $~<18$   \\
\ubma{\etaetap}   &\ubma{0.3}&\ubma{\retaetap}& \ubma{~ < \uletaetap}   & $~<27$ \\
\ubma{\etapetap}  &\ubma{0.4}&\ubma{1.7^{+4.8}_{-3.7}\pm 0.6} & \ubma{~ < 10}  & $~<47$  \\
\ubma{\etaomega}  &\ubma{4.3}&\ubma{\retaomega}& \ubma{~ < 6.2}& $~<12$  \\
\ubma{\etapomega} &\ubma{0.0}&\ubma{-0.2^{+1.3}_{-0.9}\pm 0.4}& \ubma{~ < 2.8}  & $~<60$  \\
\ubma{\etaphi}    &\ubma{0.0}&\ubma{-1.4^{+0.7}_{-0.4}\pm 0.2}& \ubma{~ < 1.0}   & $~<9$ \\
\ubma{\etapphi}   &\ubma{0.8}&\ubma{1.5^{+1.8}_{-1.5}\pm 0.4} & \ubma{~ < 4.5}  & $~<31$  \\    
\ubma{\phiphi}    &\ubma{0.3}&\ubma{\rphiphi} & \ubma{~ < 1.5} & $~<12$  \\

\dbline
\end{tabular}
\end{center}
\end{table*}

\section{Search for $B$ decays involving \azero\ mesons}
Very little is known about charmless $B$ decays with a scalar meson in
the final state.  There are also few predictions for these decays.
\cite{chernyak,minkochs}
We have searched for quasi-two-body $B$ decays with an \azero\ meson and a
pion or kaon.   This follows a previous preliminary search where evidence 
for the decay \azerompip\ was found.\cite{oldazero}  The results of the
present search are summarized in Table \ref{tab:res_azero}.  We do not confirm
the previous result which was obtained with one-quarter of this data
sample.  The difference appears to be a fluctuation.  We provide preliminary
upper limits on this and five related decay channels.  This are
the first measurements for these decays and seem to rule out the largest 
predictions for the \azeromKz\ decay from one recent paper.\cite{minkochs}

\begin{table*}[htb]
\caption{Significance ${\cal S}(\sigma)$ (including systematic uncertainties), measured branching
fraction \calB, and 90\% C.L. upper limits (UL) for $B$ decays involving
\azero\ mesons.}
\label{tab:res_azero}
\begin{center}
\begin{tabular}{lrcc}
    \dbline
Mode          &~${\cal S}(\sigma$)& ${\cal B} (10^{-6})$ & UL $(10^{-6})$  \\
\sgline
\ubma{\azerompip}&\ubma{\sazerompip~}&\ubma{\razerompip}&\ubma{<\ulazerompip} \\
\ubma{\azeromKp} &\ubma{\sazeromKp~} &\ubma{\razeromKp} &\ubma{<\ulazeromKp}  \\
\ubma{\azeromKz} &\ubma{\sazeromKz~} &\ubma{\razeromKz} &\ubma{<\ulazeromKz}  \\
\ubma{\azerozpip}&\ubma{\sazerozpip~}&\ubma{\razerozpip}&\ubma{<\ulazerozpip} \\
\ubma{\azerozKp} &\ubma{\sazerozKp~} &\ubma{\razerozKp} &\ubma{<\ulazerozKp}  \\
\ubma{\azerozKz} &\ubma{\sazerozKz~} &\ubma{\razerozKz} &\ubma{<\ulazerozKz}  \\
\dbline
\end{tabular}
\end{center}
\end{table*}

\section{Measurement of the branching fraction for the decay $B\to\Kz\pip\pim$}

We measure the branching fraction of the decay $B\to\Kz\pip\pim$.
Corrections are made for
the efficiency variation across the Dalitz plot.  From $310\pm27$ signal
events, we measure $\calB(B\to\Kz\pip\pim)=43.8\pm3.8\pm3.4\times10^{-6}$.
This is in good agreement with, but more precise than, previous results.
\cite{kspipipub}  An analysis of the Dalitz plot structure is in progress.

\section{Measurement of polarization and $CP$-violating terms in a full
angular analysis of $B\to\phi\Kstz$}
We present a full angular analysis of the decay $B\to\phi \Kstz$.  
The angular distribution of the $B\to\phi K^*$ decay 
products can be expressed in the helicity representation with
${\cal H}_i=\cos\theta_i$ and $\Phi$, where $\theta_i$ 
is the angle between the direction of one of the vector meson daughters
($i=1$ for the $K^*\to K\pi$, $i=2$ for the $\phi\to K\Kbar$)
and the direction opposite the $B$ in the resonance rest frame,
and $\Phi$ is the angle between the two resonance decay planes.
The differential decay width has three 
complex amplitudes $A_\lambda$ for the vector 
meson helicity $\lambda=0$ or $\pm 1$.\cite{bvv1,bvv2}
The decay width can be written, in terms
of $A_{\parallel}=(A_++A_-)/\sqrt{2}$, and
$A_{\perp}=(A_+-A_-)/\sqrt{2}$\,, as
\begin{eqnarray}
{{8\pi} \over {9\Gamma}}\cdot{d^3\Gamma \over 
d\calH_1 d\calH_2d\Phi} 
= {1\over|A_0|^2 + |A_{\parallel}|^2 + |A_{\perp}|^2} \times \Big[
|A_0|^2 \calH_1^2\calH_2^2
+{1\over 4}(|A_{\parallel}|^2 + |A_{\perp}|^2)
   (1-\calH_1^2)(1-\calH_2^2)~
\nonumber \\
+~{1\over 4}(|A_{\parallel}|^2 - |A_{\perp}|^2)
    (1-\calH_1^2)(1-\calH_2^2)\cos2\Phi~
 -{\rm Im}(A_{\perp}A^*_{\parallel})
    (1-\calH_1^2)(1-\calH_2^2)\sin2\Phi~
\nonumber \\
 +{\sqrt{2}}{\rm Re}(A_{\parallel}A^*_0)
   \sqrt{1-\calH_1^2}\,\calH_1
    \sqrt{1-\calH_2^2}\,\calH_2\cos\Phi~
-~{\sqrt{2}}{\rm Im}(A_{\perp}A^*_0)
    \sqrt{1-\calH_1^2}\,\calH_1
\sqrt{1-\calH_2^2}\,\calH_2\sin\Phi\Big]\,.
\nonumber 
\label{eq:helicityfull2}
\end{eqnarray}

We measure the polarization parameters 
$f_L={|A_0|^2/\Sigma|A_\lambda|^2}$,
$f_{\perp}={|A_{\perp}|^2/\Sigma|A_\lambda|^2}$,
$\phi_{\parallel} = {\rm arg}(A_{\parallel}/A_0)$,
and $\phi_{\perp} = {\rm arg}(A_{\perp}/A_0)$.
We also allow for $C\!P$-violating differences between
the $\Bbar^0$ ($Q=+1$)
and the ${B}^0$ ($Q=-1$) decay amplitudes, where
the flavor sign $Q$ is determined in the self-tagging 
final state with a $\Kbar^{*}$ or $K^{*}$:
\begin{eqnarray}
n_{\rm sig}^Q=n_{\rm sig}(1+Q{\cal A}_{C\!P})/2; ~~~~
f_{L}^{~\!Q} = f_{L}(1+Q{\cal A}_{C\!P}^{0}); ~~~~
f_{\perp}^{~\!Q} = f_{\perp}(1+Q{\cal A}_{C\!P}^{\perp});
\nonumber \\
\phi_{\parallel}^Q = \phi_{\parallel}+Q\Delta \phi_{\parallel}; ~~~~
\phi_{\perp}^Q = \phi_{\perp}+{\pi\over 2}
 + Q(\Delta \phi_{\perp}+{\pi\over 2}).~~~~~~~~~~~~~~~~~
\nonumber
\label{eq:derivedquant}
\end{eqnarray}

From the above parameters one can derive triple-product 
asymmetries ${\cal A}_T^{\parallel}$ and ${\cal A}_T^{0}$
as discussed in Ref.~\ref{bvv1}: 
\begin{eqnarray}
{\cal A}_T^{\parallel,0}= {1\over 2}\left(
{{\rm Im}(A_{\perp}A^{*}_{\parallel,0}) \over \Sigma|A_m|^2 } +
{{\rm Im}(\Abar_{\perp}\Abar^{*}_{\parallel,0}) \over \Sigma|\Abar_m|^2 }\right)\,.
\nonumber
\label{eq:asymmetryalpha4}
\end{eqnarray}

\begin{figure}[hb!]
\begin{center}
 \scalebox{0.9}{
  \includegraphics[width=.4\linewidth]{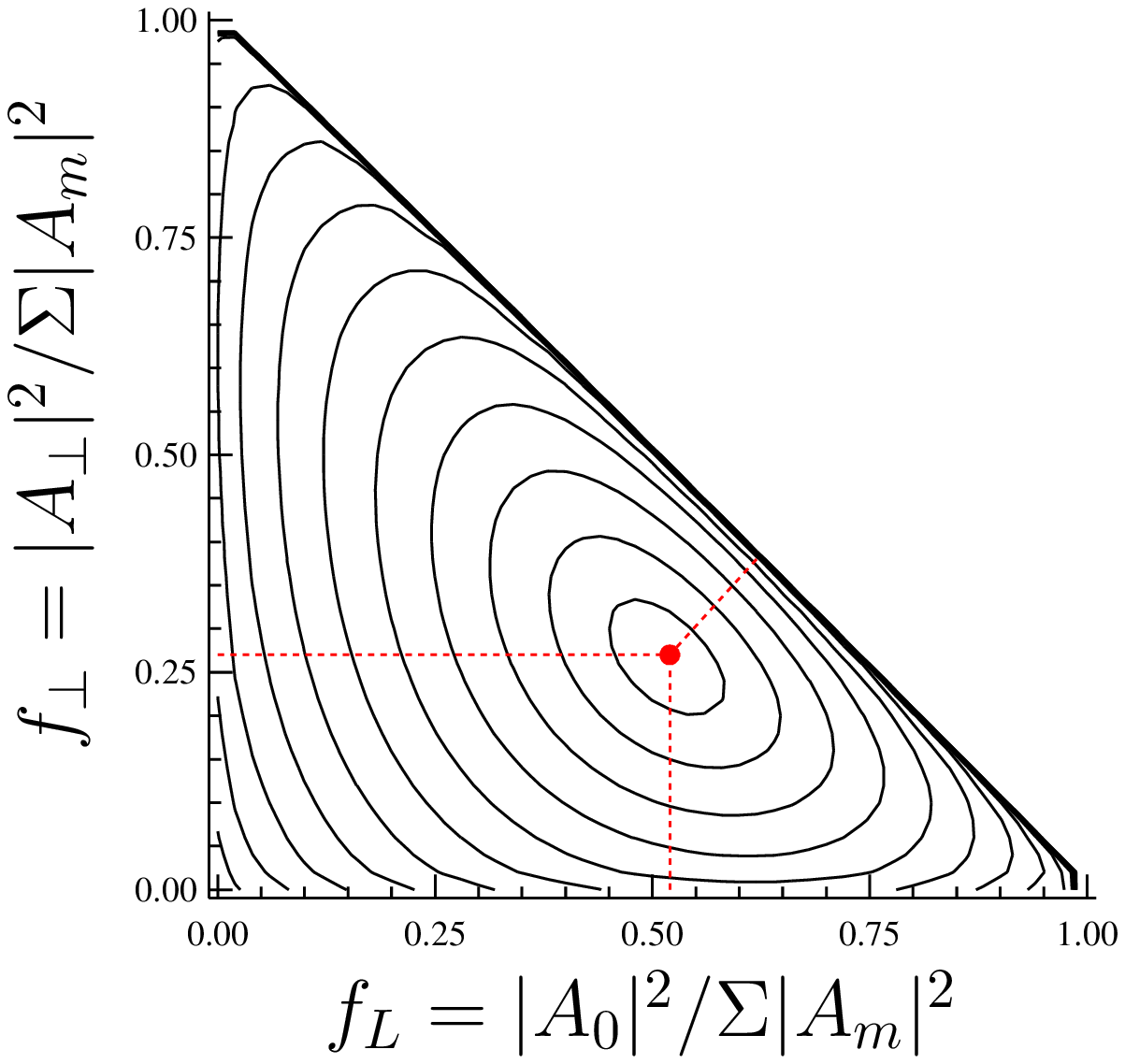}\hspace{0.8cm}
  \includegraphics[width=.4\linewidth]{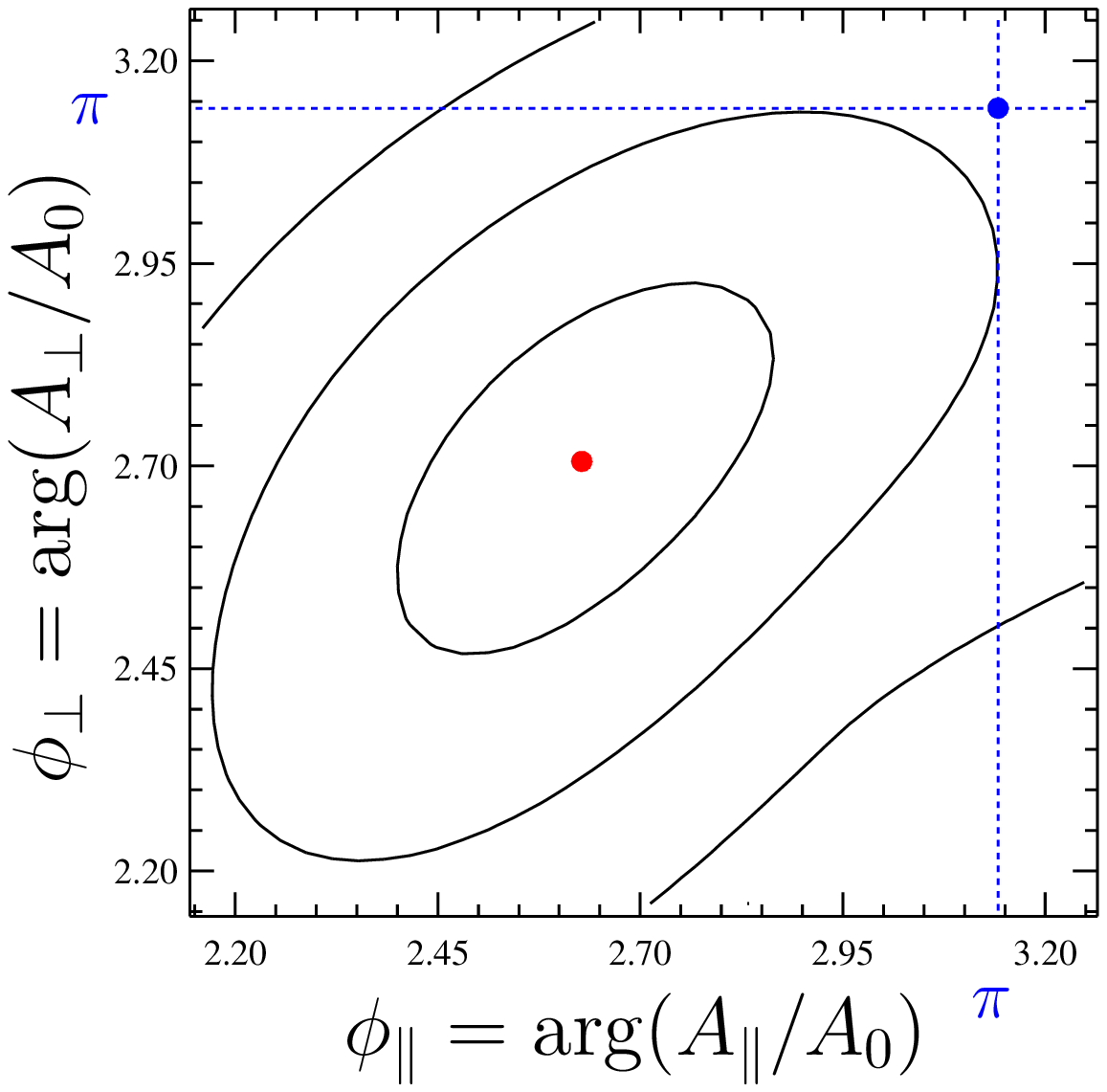} }
 \vspace*{3mm}
 \scalebox{0.9}{\hfill (a)\hspace{7cm}(b)\hfill}
 \scalebox{0.9}{
  \includegraphics[width=.4\linewidth]{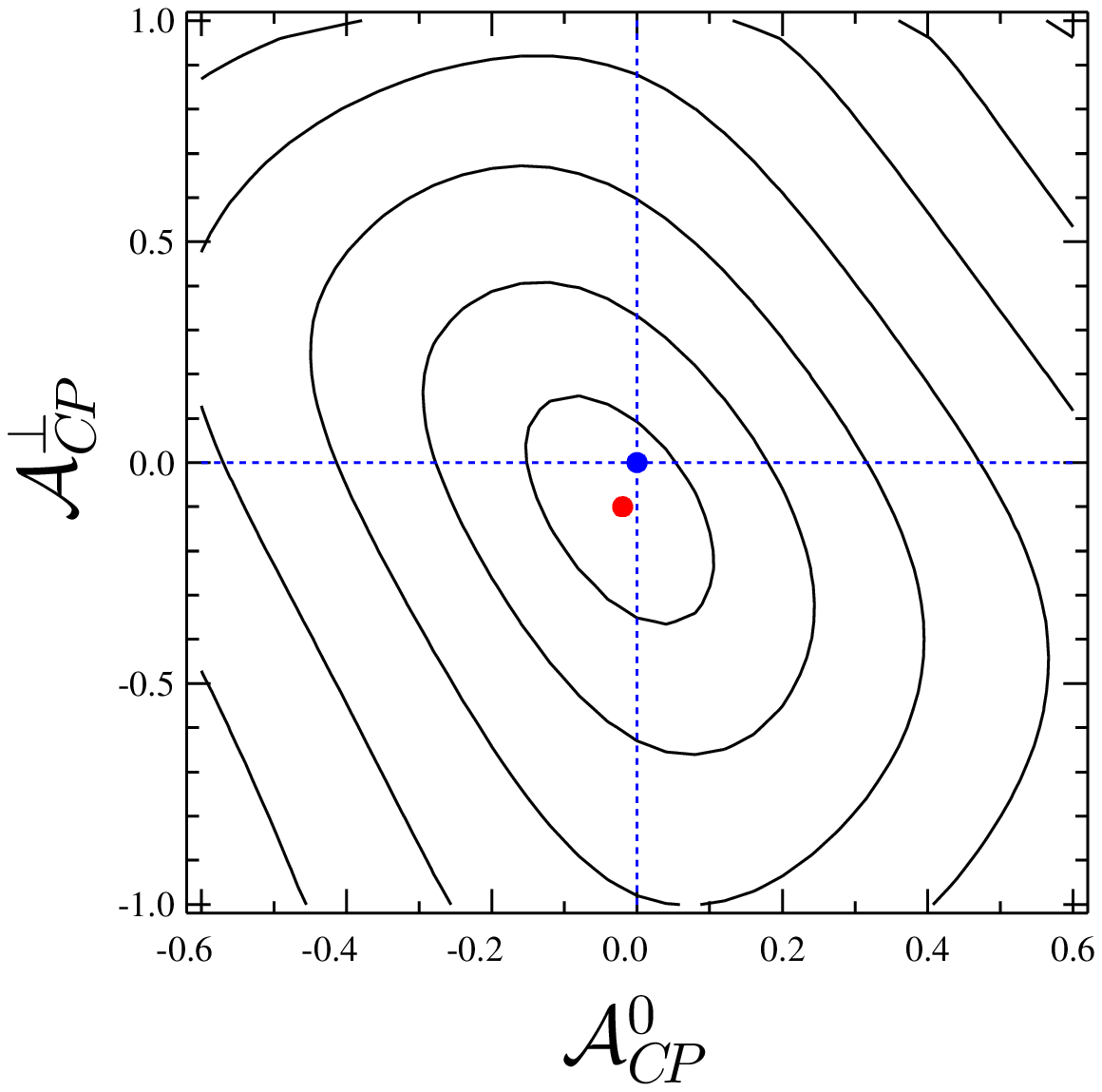}\hspace{1cm}
  \includegraphics[width=.4\linewidth]{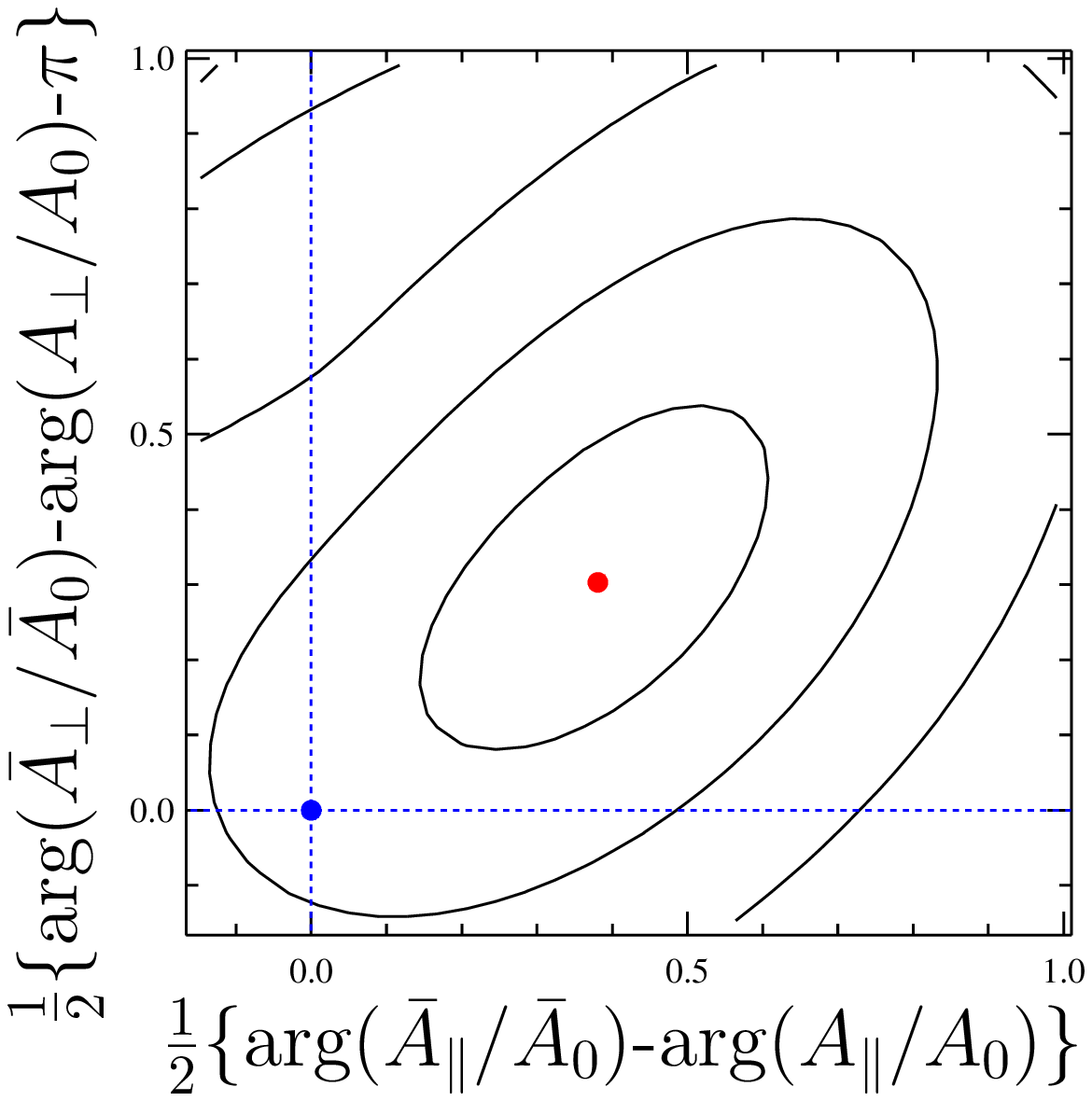} }
 \scalebox{0.9}{\hfill (c)\hspace{7cm}(d)\hfill}
\end{center}
 \caption{\label{fig:phikst}Contour plots with 1$\sigma$ intervals
derived from the fit $-2\ln{\cal L}$ distributions for (a) polarization
fractions $f_\perp$ and $f_L$, (b) $CP$-even and $CP$-odd transverse phases
([$\pi,\pi$] point expected if no final-state interactions), (c) 
asymmetry parameters sensitive to direct $CP$ violation; (d) phases
of the triple-product asymmetries that are sensitive to new physics.
\protect\cite{bvv1}}
\end{figure}

The longitudinal polarization in this decay is found to be $0.52\pm0.07\pm0.02$ 
as seen in Table \ref{tab:phikst} and Fig.~\ref{fig:phikst}(a); this
value is surprising since naive expectations and measurements for
$B\to\rho\rho$ indicate a value very close to 1.  This confirms earlier
measurements by \babar\ \cite{BABARphikst} and Belle \cite{Bellephikst} and is 
still not understood theoretically.
Also shown in Fig.~\ref{fig:phikst}(b)-(d) are measurements involving the
other quantities determined in the fit.

\begin{table}[btp]
\caption{
We show results for the ten primary signal fit parameters
and the secondary triple-product asymmetries.
All results include systematic errors quoted last.
The dominant correlations coefficients are also shown.
}
\label{tab:phikst}
\begin{center}
\begin{tabular}{cclccl}
\dbline
   Fit param. & Fit result & Corr. &
   Fit param. & Fit result & Corr. \cr
\sgline
  $n_{\rm sig}$ (events) & $129\pm 14\pm 9$ & &
  ${\cal A}_{C\!P}$ & $-0.12\pm{0.10}\pm 0.03$  & 
\cr
  ${f_L}$ & $0.52\pm{0.07}\pm 0.02$  & \multirow{2}{13mm}{{\Large\}}~$-52\%$} &
  ${\cal A}_{C\!P}^0$ & $-0.02\pm{0.12}\pm 0.01$ & \multirow{2}{13mm}{{\Large\}}~$-52\%$}
\cr
  ${f_\perp}$ & $0.27\pm{0.07}\pm 0.02$  &  &
  ${\cal A}_{C\!P}^{\perp}$ & $-0.10^{+0.25}_{-0.27}\pm 0.04$  &  
\cr
  ${\phi_\parallel}$ (rad) & $2.63^{+0.24}_{-0.23}\pm 0.04$ &  \multirow{2}{13mm}{{\Large\}}~$+59\%$} &
  $\Delta \phi_{\parallel}$ (rad) & $0.38^{+0.23}_{-0.24}\pm 0.04$ & \multirow{2}{13mm}{{\Large\}}~$+59\%$}
\cr
  ${\phi_\perp}$ (rad) & $2.71^{+0.22}_{-0.24}\pm 0.03$  &  &
  $\Delta \phi_{\perp}$ (rad) & $0.30^{+0.24}_{-0.22}\pm 0.03$   & 
\cr
\sgline
  ${\cal A}_T^{\parallel}$ & $+0.02\pm{0.05}\pm 0.01$  &  &
  ${\cal A}_T^{0}$ & $+0.11\pm{0.07}\pm 0.01$  &  
\cr
\dbline
\end{tabular}
\end{center}
\end{table}

\section*{Acknowledgments}
I would like to thank the organizers for an enjoyable and stimulating conference.
I thank my \babar\ colleagues for their assistance and helpful discussions, 
especially F. Blanc, W. Ford, A. Gritsan, A. H\"ocker, C. Lee, and F. Palombo.
I also gratefully acknowledge useful discussions with M. Gronau and 
Y. Grossman.

\end{document}

%% file: Definitions.tex
\RequirePackage{xspace}

\newcommand{\calH}{\ensuremath{{\cal H}}}
\newcommand{\pvec}{{\bf p}}

\newcommand{\acp}{\ensuremath{\calA_{ch}}}
\newcommand{\calB}{\ensuremath{{\cal B}}}

\newcommand{\bfemsix}{${\cal B}(10^{-6}$)}

\newcommand{\DE}{\ensuremath{\Delta E}}

\newcommand\etal{{\it et al.}}
\newcommand{\half}{\ensuremath{{1\over2}}}

\newcommand{\bfig}{\begin{figure}[htbpc!]}
\newcommand{\efig}{\end{figure}}
\newcommand\bef{\begin{figure}}
\newcommand\edf{\end{figure}}
\newcommand\dbline{\noalign{\vskip 0.10truecm\hrule}\noalign{\vskip 2pt}\noalign{\hrule\vskip 0.10truecm}}
\providecommand{\tbline}{\noalign{\vskip 0.05truecm\hrule\vskip0.05truecm}}
\newcommand\sgline{\noalign{\vskip 0.10truecm\hrule\vskip 0.10truecm}}
\newcommand\beq{\begin{equation}}
\newcommand\eeq{\end{equation}}
\newcommand\bear{\begin{array}}
\newcommand\enar{\end{array}}
\newcommand\beqa{\begin{eqnarray}}
\newcommand\eeqa{\end{eqnarray}}
\newcommand\ben{\begin{enumerate}}
\newcommand\een{\end{enumerate}}

\newcommand{\UfourS}{\ensuremath{\Upsilon(4S)}}

\newcommand{\etaprp}{\ensuremath{\eta^{(\prime)}}}		

\newcommand{\Kstz}{\ensuremath{\Kstarz}}

   \newcommand{\rhop}{\ensuremath{\rho^+}}
   
   \newcommand{\rhom}{\ensuremath{\rho^-}}
   
\newcommand{\fetapip}{\ensuremath{\eta\pi^+}}
\newcommand{\etapip}{\ensuremath{\Bp\ra\fetapip}}

\newcommand{\retapip}{\ensuremath{xx^{+xx}_{-xx}\pm xx}}

\newcommand{\Aetapip}{\ensuremath{xx^{+xx}_{-xx}\pm xx}}

\newcommand{\setapip}{\ensuremath{xx}}

\newcommand{\fetaKp}{\ensuremath{\eta K^+}}
\newcommand{\etaKp}{\ensuremath{\Bp\ra\fetaKp}}

\newcommand{\retaKp}{\ensuremath{xx^{+xx}_{-xx}\pm xx}}

\newcommand{\AetaKp}{\ensuremath{xx^{+xx}_{-xx}\pm xx}}

\newcommand{\setaKp}{\ensuremath{xx}}

\newcommand{\fetaKz}{\ensuremath{\eta\Kz}}
\newcommand{\etaKz}{\ensuremath{\Bz\ra\fetaKz}}

\newcommand{\retaKz}{\ensuremath{xx^{+xx}_{-xx}\pm xx}}

\newcommand{\uletaKz}{\ensuremath{xx}}

\newcommand{\setaKz}{\ensuremath{xx}}

\newcommand{\fetapiz}{\ensuremath{\eta\piz}\xspace}
\newcommand{\etapiz}{\ensuremath{\Bz\ra\fetapiz}\xspace}

\newcommand{\retapiz}{\ensuremath{xx^{+xx}_{-xx}\pm xx}\xspace}

\newcommand{\uletapiz}{\ensuremath{xx}\xspace}

\newcommand{\setapiz}{\ensuremath{xx}\xspace}

\newcommand{\fetaKst}{\ensuremath{\eta K^{*}}}
\newcommand{\etaKst}{\ensuremath{\B\ra\fetaKst}}

\newcommand{\fetaKstp}{\ensuremath{\eta K^{*+}}}
\newcommand{\etaKstp}{\ensuremath{\Bp\ra\fetaKstp}}

\newcommand{\retaKstp}{\ensuremath{xx^{+xx}_{-xx}\pm xx}}

\newcommand{\AetaKstp}{\ensuremath{xx\pm xx\pm xx}}
\newcommand{\setaKstp}{\ensuremath{xx}}

\newcommand{\fetaKstz}{\ensuremath{\eta K^{*0}}}
\newcommand{\etaKstz}{\ensuremath{\Bz\ra\fetaKstz}}

\newcommand{\retaKstz}{\ensuremath{xx^{+xx}_{-xx}\pm xx}}

\newcommand{\AetaKstz}{\ensuremath{xx\pm xx \pm xx}}
\newcommand{\setaKstz}{\ensuremath{xx}}

\newcommand{\fetarhop}{\ensuremath{\eta\rho^+}}
\newcommand{\etarhop}{\ensuremath{\Bp\ra\fetarhop}}

\newcommand{\retarhop}{\ensuremath{xx^{+xx}_{-xx}\pm xx}}

\newcommand{\uletarhop}{\ensuremath{xx}}

\newcommand{\setarhop}{\ensuremath{xx}}

\newcommand{\fetarhoz}{\ensuremath{\eta\rho^0}}
\newcommand{\etarhoz}{\ensuremath{\Bz\ra\fetarhoz}}

\newcommand{\retarhoz}{\ensuremath{xx^{+xx}_{-xx}\pm xx}}

\newcommand{\uletarhoz}{\ensuremath{xx}}

\newcommand{\setarhoz}{\ensuremath{xx}}

\newcommand{\fetappip}{\ensuremath{\etapr\pip}}
\newcommand{\etappip}{\ensuremath{\Bp\ra\fetappip}}

\newcommand{\retappip}{\ensuremath{xx^{+xx}_{-xx} \pm xx}}

\newcommand{\uletappip}{\ensuremath{xx}}

\newcommand{\setappip}{\ensuremath{xx}}

\newcommand{\fetapKp}{\ensuremath{\etapr K^+}}
\newcommand{\etapKp}{\ensuremath{\Bp\ra\fetapKp}}

\newcommand{\fetapKz}{\ensuremath{\etapr K^0}}
\newcommand{\etapKz}{\ensuremath{\Bz\ra\fetapKz}}

\newcommand{\fetappiz}{\ensuremath{\etapr\piz}\xspace}
\newcommand{\etappiz}{\ensuremath{\Bz\ra\fetappiz}\xspace}

\newcommand{\retappiz}{\ensuremath{xx^{+xx}_{-xx} \pm xx}\xspace}

\newcommand{\uletappiz}{\ensuremath{xx}\xspace}

\newcommand{\setappiz}{\ensuremath{xx}\xspace}

\newcommand{\etapKst}{\ensuremath{\B\ra\etapr K^{*}}}

\newcommand{\etapKstp}{\ensuremath{\Bp\ra\etapr K^{*+}}}

\newcommand{\retapKstp}{\ensuremath{xx^{+xx}_{-xx}\pm xx}}

\newcommand{\uletapKstp}{\ensuremath{xx}}

\newcommand{\setapKstp}{\ensuremath{xx}}

\newcommand{\fetapKstz}{\ensuremath{\etapr K^{*0}}}
\newcommand{\etapKstz}{\ensuremath{\Bz\ra\fetapKstz}}

\newcommand{\retapKstz}{\ensuremath{xx^{+xx}_{-xx}\pm xx}}

\newcommand{\uletapKstz}{\ensuremath{xx}}

\newcommand{\setapKstz}{\ensuremath{xx}}

\newcommand{\fetaprhop}{\ensuremath{\etapr\rho^+}}
\newcommand{\etaprhop}{\ensuremath{\Bp\ra\fetaprhop}}

\newcommand{\retaprhop}{\ensuremath{xx^{+xx}_{-xx}\pm xx}}

\newcommand{\uletaprhop}{\ensuremath{xx}}

\newcommand{\setaprhop}{\ensuremath{xx}}

\newcommand{\fetaprhoz}{\ensuremath{\etapr\rho^0}}
\newcommand{\etaprhoz}{\ensuremath{\Bz\ra\fetaprhoz}}

\newcommand{\retaprhoz}{\ensuremath{xx^{+xx}_{-xx}\pm xx}}

\newcommand{\uletaprhoz}{\ensuremath{xx}}

\newcommand{\setaprhoz}{\ensuremath{xx}}

\newcommand{\fomegapiz}{\ensuremath{\omega\pi^0}\xspace}
\newcommand{\omegapiz}{\ensuremath{\Bz\ra\fomegapiz}\xspace}

\newcommand{\romegapiz}{\ensuremath{xx\pm xx \pm xx}\xspace}

\newcommand{\ulomegapiz}{\ensuremath{xx}\xspace}

\newcommand{\azero}{\ensuremath{a_0}}

\newcommand{\fazeromKz}{\ensuremath{a_0^- K^0}}
\newcommand{\fazerozKz}{\ensuremath{a_0^0 K^0}}

\newcommand{\fazeromKp}{\ensuremath{a_0^- K^+}}
\newcommand{\fazerozKp}{\ensuremath{a_0^0 K^+}}
\newcommand{\fazerompip}{\ensuremath{a_0^-\pi^+}}
\newcommand{\fazerozpip}{\ensuremath{a_0^0\pi^+}}

\newcommand{\azeromKz}{\ensuremath{\Bm\ra\fazeromKz}}
\newcommand{\azerozKz}{\ensuremath{\Bz\ra\fazerozKz}}

\newcommand{\azeromKp}{\ensuremath{\Bz\ra\fazeromKp}}
\newcommand{\azerozKp}{\ensuremath{\Bp\ra\fazerozKp}}
\newcommand{\azerompip}{\ensuremath{\Bz\ra\fazerompip}}
\newcommand{\azerozpip}{\ensuremath{\Bp\ra\fazerozpip}}

\newcommand{\razeromKz}{\ensuremath{xx^{+xx}_{-xx}\pm xx}}

\newcommand{\ulazeromKz}{\ensuremath{xx}\xspace}

\newcommand{\sazeromKz}{\ensuremath{xx}\xspace}

\newcommand{\razerozKz}{\ensuremath{xx^{+xx}_{-xx}\pm xx}}

\newcommand{\ulazerozKz}{\ensuremath{xx}\xspace}

\newcommand{\sazerozKz}{\ensuremath{xx}\xspace}

\newcommand{\razeromKp}{\ensuremath{xx^{+xx}_{-xx}\pm xx}}

\newcommand{\ulazeromKp}{\ensuremath{xx}\xspace}

\newcommand{\sazeromKp}{\ensuremath{xx}\xspace}

\newcommand{\razerozKp}{\ensuremath{xx^{+xx}_{-xx}\pm xx}}

\newcommand{\ulazerozKp}{\ensuremath{xx}\xspace}

\newcommand{\sazerozKp}{\ensuremath{xx}\xspace}

\newcommand{\razerompip}{\ensuremath{xx^{+xx}_{-xx}\pm xx}}

\newcommand{\ulazerompip}{\ensuremath{xx}\xspace}

\newcommand{\sazerompip}{\ensuremath{xx}\xspace}

\newcommand{\razerozpip}{\ensuremath{xx^{+xx}_{-xx}\pm xx}}

\newcommand{\ulazerozpip}{\ensuremath{xx}\xspace}

\newcommand{\sazerozpip}{\ensuremath{xx}\xspace}

\newcommand{\fphipiz}{\ensuremath{\phi\pi^0}\xspace}
\newcommand{\phipiz}{\ensuremath{\Bz\ra\fphipiz}\xspace}

\newcommand{\rphipiz}{\ensuremath{xx\pm xx \pm xx}\xspace}

\newcommand{\ulphipiz}{\ensuremath{xx}\xspace}

\newcommand{\sphipiz}{\ensuremath{xx}\xspace}